\documentclass[hyper]{JHEP} 

\usepackage{epsfig}

\newcommand\fverb{\setbox\pippobox=\hbox\bgroup\verb}

\newcommand\fverbdo{\egroup\medskip\noindent%

            \fbox{\unhbox\pippobox}\ }

\newcommand\fverbit{\egroup\item[\fbox{\unhbox\pippobox}]}

\newbox\pippobox

\title{Null Dimensional Reduction of M2-Brane}
\author{J. Kluso\v{n}\\
Department of
Theoretical Physics and Astrophysics\\
Faculty of Science, Masaryk University\\
Kotl\'{a}\v{r}sk\'{a} 2, 611 37, Brno\\
Czech Republic\\
E-mail: \email{klu@physics.muni.cz}} \preprint{}

 \abstract{We perform a general reduction of an M2-brane on a space-time that admits a null Killing vector leading to fundamental string and D2-brane action in 
 Newton-Cartan background.}

\def\hv{\hat{v}}

\def\hH{\hat{H}}
\def\hV{\hat{V}}

\def\bA{\mathbf{A}}

\def\balpha{\bar{\alpha}}
\def\bbeta{\bar{\beta}}
\def\bgamma{\bar{\gamma}}
\def\bdelta{\bar{\delta}}

\def\bg{\bar{g}}

\def\be{\begin{equation}}

\def\ee{\end{equation}}

\def\bea{\begin{eqnarray}}

\def\eea{\end{eqnarray}}

\def\mH{\mathcal{H}}

\newcommand{\hg}{\hat{g}}

\def \bA{\mathbf{A}}

\newcommand{\mL}{\mathcal{L}}

\begin{document}
\section{Introduction and Summary}
Idea of non-relativistic string theory was firstly proposed twenty years
ago in two papers \cite{Gomis:2000bd,Danielsson:2000gi}. These works studied 
strings with non-relativistic spectrum in flat space-time and they were related
to the analysis of strings in the background with critical electric fields
\cite{Seiberg:2000ms} which is related to the famous relation between string theory 
and noncommutative geometry \cite{Seiberg:1999vs}. The  generalization from flat to special curved background was performed in \cite{Gomis:2005pg}. New interest in non-relativistic string theories begun recently in the context of AdS/CFT correspondence 
\cite{Christensen:2013lma,Christensen:2013rfa} where torsional Newton-Cartan (NC) geometry
was firstly observed as the geometry of boundary theory. More importantly, this torsional
NC geometry is also consistent background for non-relativistic strings \cite{Harmark:2017rpg,Harmark:2019upf,Harmark:2018cdl}
\footnote{For related works, see for example \cite{Bergshoeff:2021bmc,Fontanella:2021hcb,Kluson:2021sym,Hansen:2020wqw,Harmark:2020vll,Hansen:2020pqs,Gomis:2020fui,Kluson:2020aoq,Kluson:2019xuo,Blair:2019qwi,Bergshoeff:2019pij,Kluson:2019avy,Gomis:2019zyu,Kluson:2019uza,Hansen:2018ofj,Kluson:2018vfd,Kluson:2018grx,Bergshoeff:2018yvt,Kluson:2018egd}.}. In fact,there are basically two ways how to construct non-relativistic string theory. The first one is based on specific non-relativistic limit 
\cite{Andringa:2012uz,Bergshoeff:2018yvt,Bergshoeff:2019pij} while the second one on null reduction of
the world-sheet string in arbitrary general background with null isometry
\cite{Harmark:2017rpg,Harmark:2020vll,Harmark:2019upf,Harmark:2018cdl}. It was then shown in \cite{Harmark:2019upf} how these two non-relativistic string theories are related. In fact, it was also shown in \cite{Kluson:2018egd} that the non-relativistic string in torisonal Newton-Cartan background is defined by specific T-duality transformations along direction with null isometry. More precisely, we start with the relativistic string in the background with light-like isometry. Then performing T-duality along this direction we obtain string in torsional Newton-Cartan background \cite{Harmark:2017rpg,Harmark:2020vll,Harmark:2019upf,Harmark:2018cdl}, see also \cite{Kluson:2019xuo,Kluson:2019avy}.

In this paper we would like to study the question whether it is possible to define non-relativistic string from M2-brane through double dimensional reduction along null isometry of the M-theory background. It is well known that dimensional reduction of M-theory along spatial circle leads to Type IIA theory where M2-brane maps either into fundamental string or to D2-brane \cite{Townsend:1995af,Duff:1992hu}, for excelent review, see \cite{Townsend:1996xj}. In this paper we proceed in similar way when we analyse M2-brane in the M-theory background with null isometry. This is very important problem which, as far as we know, has not been studied yet, with exception of recent paper \cite{Lambert:2020scy}
\footnote{Of course there is well known procedure of DLCQ of M-theory 
\cite{Banks:1996vh,Sen:1997we,Seiberg:1997ad}.It would be certainly very interesting to see how our approach is related to the Matrix theory description of M-theory.}.
We firstly consider M2-brane extended along null isometry direction and perform its double dimensional reduction. In order to perform this procedure we write M2-brane
using auxiliary three dimensional metric and presume that this metric has null isometry as well. As a result we obtain Polyakov like form of non-relativistic string 
with Newton-Cartan world-sheet metric so that it can be interpreted as a string with non-relativistic world-sheet. However when we eliminate components of the world-sheet Newton-Cartan gravity by solving their equations of motion we obtain standard square root form of the string action with the induced volume element corresponding to the target space-time Newton-Cartan gravity. We also find that for given
eleven dimensional background with null isometry there is a family of Newton-Cartan backgrounds that are related by simple redefinitions which is similar result as in
\cite{Bergshoeff:2018yvt}.

As the next step we consider M2-brane that is transverse to the null isometry direction. Following similar analysis as in case of spatial dimensional reduction
\cite{Townsend:1995af} we find that this action corresponds to D2-brane action 
in Newton-Cartan background.

In summary, we consider null dimensional reduction of M-theory background and we find that M2-brane maps either to fundamental string or D2-brane in Newton-Cartan background. We find that these actions basically contain pull back of the volume element of target space Newton-Cartan background which is nice result that  implies many new questions. The first one is how these non-relativistic strings and D2-brane actions are related to the non-relativistic string actions that were found in previous works. This question is closely related to the problem how null dimensional reduction of M-theory background is related to corresponding stringy Newton-Cartan background studied recently in 
\cite{Bergshoeff:2021bmc}. And finally it would be certainly interesting to see how
our work could be related to Matrix theory conjecture. We hope to return to these problems in future. 

This paper is organized as follows. In the next section (\ref{second}) we review double dimensional reduction of M2-brane along spatial dimensional direction of M-theory background. Then in section (\ref{third}) we perform null dimensional reduction of M-theory and double dimensional reduction of M2-brane. In section (\ref{fourth}) we discuss Hamiltonian formulation of this non-relativistic string. Finally in section (\ref{fifth}) we perform transverse dimensional reduction of M2-brane that leads to D2-brane in non-relativistic background.

\section{Review of Double Dimensional Reduction of M2-brane along Spatial Circle}\label{second}
In this section we perform double dimensional reduction of M2-brane that leads to fundamental string in Type IIA theory. This procedure is well known when M2-brane is defined using the square root form of the action (See for example excellent review \cite{Townsend:1996xj}). However we would lie to perform this analysis with the help of Polyakov form of the M2-brane action
\footnote{
For simplicity we restrict ourselves to the case of vanishing three form field in M-theory 
keeping in mind that generalization of the analysis with inclusion of this field is straightforward.}. 

In order to perform dimensional reduction we consider eleven dimensional background 
metric in the form \cite{Townsend:1996xj}
\begin{equation}\label{lineM}
ds^2=\hg_{MN}dx^Mdx^N=e^{-\frac{2}{3}\phi}g_{\mu\nu}dx^\mu dx^\nu+
e^{\frac{4}{3}\phi}(dy-C_\mu dx^\mu)^2 \ ,
\end{equation}
where $\mu,\nu=0,1,\dots,9$ and where $y$ is compact direction. 
As is well known natural probe of M-theory is M2-brane which is three dimensional extended
object with the action 
\begin{equation}\label{SM2}
S=-T_{M2}\int d^3\xi \sqrt{-\det \hg_{\bar{\alpha}
		\bar{\beta}}} \ , \quad \hg_{\bar{\alpha}\bar{\beta}}=
\hg_{MN}\partial_{\bar{\alpha}}x^M\partial_{\bar{\beta}}x^N \ , 
\end{equation}
where $T_{M2}$ is M2-brane tension, $\xi^{\balpha},\balpha=0,1,2$ are  coordinates that label three dimensional world-volume of M2-brane. Further, $x^M, M,N=0,1,\dots,10$ label embedding of M2-brane into target space-time. 

Let us now consider M2-brane that wraps compact direction labelled by $y$. Standard procedure is to use double dimensional reduction in the action (\ref{SM2})  \cite{Townsend:1996xj}. However for letter purposes we proceed with slightly different way when we consider form of M2-brane action written using three dimensional auxiliary metric $\gamma_{\balpha\bbeta}$
so that we can rewrite the action (\ref{SM2}) into the form
\begin{equation}\label{M2pol}
S=-\frac{T_{M2}}{2}\int d^3\xi \sqrt{-\hat{\gamma}} (\hat{\gamma}^{\bar{\alpha}\bar{\beta}}\hg_{\bar{\beta}\bar{\alpha}}-1) \ .
\end{equation}
In order to see an equivalence between this Polyakov-like M2-brane action 
(\ref{M2pol}) and (\ref{SM2}) let us consider equation of motion for $\gamma^{\balpha\bbeta}$
that follow from (\ref{M2pol})
\begin{equation}
-\frac{1}{2}\sqrt{-\gamma}\gamma_{\balpha\bbeta}(\gamma^{\bgamma\bdelta}\hg_{\bdelta\bgamma}-1)+
\sqrt{-\gamma}\bg_{\balpha\bbeta}=0 \ .
\end{equation}
This equation can be solved for $\gamma_{\balpha\bbeta}$ as  $\gamma_{\balpha\bbeta}=g_{\balpha\bbeta}$. Then inserting this result into (\ref{M2pol}) we obtain an action (\ref{SM2}).

As we wrote above we presume that M2-brane wraps $y-$direction so that we can identify $\xi^2$ with $y$. As a result we get  following components of induced metric
\begin{equation}\label{hgin}
 \hg_{22}=e^{\frac{4}{3}\phi} \ , \quad  \hg_{2\beta}=-e^{\frac{4}{3}\phi}C_\mu\partial_\beta x^\mu \ , \quad 
\hg_{\alpha 2}=-e^{\frac{4}{3}\phi}\partial_\alpha x^\mu C_\mu \ , \quad 
\hg_{\alpha\beta}=e^{-2\phi}g_{\alpha\beta}+e^{\frac{4}{3}\phi}
C_\alpha C_\beta \ ,
\end{equation}
where $\alpha,\beta=0,1$.
Let us insert this ansatz into (\ref{SM2}) and we obtain 
\begin{eqnarray}\label{SMred}
& &S=-T_{M2}\int dy d^2\xi\sqrt{-\det \hg_{\bar{\alpha}\bar{\beta}}}=\nonumber \\
& &=-T_{M2}\int dy \int d^2\xi\sqrt{-\det \left(\hg_{\alpha\beta}-\hg_{\alpha y}\frac{1}{\hg_{yy}}
		\hg_{y\beta}\right)\hg_{yy}}=T_{FS}\int d^2\xi\sqrt{-\det g_{\alpha\beta}}
	\nonumber \\
\end{eqnarray}
which is the standard  Nambu-Goto action for fundamental string in Type IIA theory when 
we performed identification between string tension $T_{FS}$ and M2-brane tension in the form
\begin{equation}
T_{FS}=T_{M2}\int dy \ . 
\end{equation}
Let us now perform the same double dimensional reduction in case
of the Polyakov form of M2-brane action. Inserting the ansatz 
 (\ref{hgin}) into (\ref{M2pol}) we obtain
\begin{equation}\label{M2polin}
S=-\frac{T_{M2}}{2}\int d^3\xi
\sqrt{-\det \hat{\gamma}}
(\hat{\gamma}^{\alpha\beta}\hg_{\beta\alpha}+2\hat{\gamma}^{y\alpha}\hg_\alpha
+\hat{\gamma}^{22}\hat{g}_{yy}-1) \ . 
\end{equation}
Since we presume that M2-brane reduces in the same way as the background metric
it is natural to consider the same ansatz for induced metric as in (\ref{lineM}).
In more details, let us presume that the world-volume metric has the form
\begin{equation}\label{hgamma}
\hat{\gamma}_{yy}=e^{\frac{4}{3}\varphi} \ , \quad \hat{\gamma}_{u\alpha}=-e^{\frac{4}{3}\varphi}c_\alpha \ , \quad  \hat{\gamma}_{\alpha\beta}=e^{-\frac{2}{3}\varphi}\gamma_{\alpha\beta}+e^{\frac{4}{3}\varphi}c_\alpha c_\beta \ . 
\end{equation}
It is easy to see that 
\begin{equation}
\det \hat{\gamma}_{\hat{\alpha}\hat{\beta}}=\det \gamma_{\alpha\beta} \ . 
\end{equation}
Further, the metric inverse to (\ref{hgamma}) has the form 
\begin{equation}
\hat{\gamma}^{\alpha\beta}=e^{\frac{2}{3}\varphi}\gamma^{\alpha\beta} \ , \quad 
\hat{\gamma}^{y\beta}=e^{\frac{2}{3}\varphi}c_\gamma \gamma^{\gamma\beta} \ , \quad 
\hat{\gamma}^{\alpha y}=e^{\frac{2}{3}\varphi}\gamma^{\alpha\delta}c_\delta 
\ ,  \quad 
\hat{\gamma}^{yy}=e^{-\frac{4}{3}\varphi}+e^{\frac{2}{3}}c_\alpha \gamma^{\alpha\beta}c_\beta  \ . 
\end{equation}
Then the action (\ref{M2polin}) is equal to
\begin{eqnarray}\label{Sext}
& &S=-\frac{T_{M2}}{2}\int d^3\xi \sqrt{-\gamma}
(e^{\frac{2}{3}\varphi}\gamma^{\alpha\beta}(e^{-\frac{2}{3}\phi}g_{\beta\alpha}+e^{\frac{4}{3}\phi}C_{\beta}C_\alpha)-\nonumber \\
& &-2 e^{\frac{2}{3}\varphi}c_\gamma \gamma^{\gamma\alpha}C_\alpha e^{\frac{4}{3}\phi}
+(e^{-\frac{4}{3}\varphi}+e^{\frac{2}{3}\varphi}c_\alpha \gamma^{\alpha\beta}c_\beta)e^{\frac{4}{3}\phi}-1) \ . 
\nonumber \\
\end{eqnarray}
In order to see an equivalence of the action (\ref{Sext})  with  (\ref{SMred})  let us solve the equation of motion for $c_\alpha$ that follow from (\ref{Sext})
\begin{equation}
-e^{\frac{2}{3}(2\phi+\varphi)}\gamma^{\alpha\beta}C_\beta+
e^{\frac{2}{3}(2\phi+\varphi)}\gamma^{\alpha\beta}c_\beta=0
\end{equation}
that has solution $c_\alpha=C_\alpha$. Further, equation of motion for $\varphi$  has the form
\begin{eqnarray}
& &e^{\frac{2}{3}\varphi}\gamma^{\alpha\beta}(e^{-\frac{2}{3}\phi}g_{\beta\alpha}+e^{\frac{4}{3}\phi}C_{\beta}C_\alpha)
-\nonumber \\
& &-2e^{\frac{2}{3}\varphi}c_\alpha \gamma^{\alpha\beta}C_\beta e^{\frac{4}{3}\phi}
-2e^{-\frac{4}{3}\varphi}e^{\frac{4}{3}\phi}
+e^{\frac{2}{3}\varphi}c_\alpha \gamma^{\alpha\beta}c_\beta e^{\frac{4}{3}\phi}=0  \ .
\nonumber \\
\end{eqnarray}
Using $c_\alpha=C_\alpha$ the previous equation reduces into
\begin{equation}\label{eqphi}
e^{\frac{2}{3}(\varphi-\phi)}\gamma^{\alpha\beta}g_{\beta\alpha}
-2 e^{\frac{4}{3}(\phi-\varphi)}=0 \ . 
	\end{equation}
Finally,  the equation of motion for $\gamma^{\alpha\beta}$ has the form
\begin{eqnarray}
& &-\frac{1}{2}\gamma_{\beta\alpha}	
(e^{\frac{2}{3}\varphi}\gamma^{\gamma\delta}(e^{-\frac{2}{3}\phi}g_{\delta\gamma}+e^{\frac{4}{3}\phi}C_{\gamma}C_\delta)-\nonumber \\
& &-2 e^{\frac{2}{3}\varphi}c_\gamma \gamma^{\gamma\alpha}C_\alpha e^{\frac{4}{3}\phi}
	+(e^{-\frac{4}{3}\varphi}+e^{\frac{2}{3}\varphi}c_\alpha \gamma^{\alpha\beta}c_\beta)e^{\frac{4}{3}\phi}-1)+\nonumber \\
& &+e^{\frac{2}{3}(\varphi-\phi)}g_{\beta\alpha}+
e^{\frac{2}{3}\varphi+\frac{4}{3}\phi}C_\beta C_\alpha
-2 e^{\frac{2}{3}\varphi+\frac{4}{3}\phi}c_\beta C_\alpha
+e^{\frac{2}{3}\varphi+\frac{4}{3}\phi}c_\beta c_\alpha=0
\nonumber \\
\end{eqnarray}
that for $c_\alpha=C_\alpha$ simplifies as
\begin{eqnarray}\label{eqgammasim}
-\frac{1}{2}\gamma_{\beta\alpha}	
(e^{\frac{2}{3}(\varphi-\phi)}\gamma^{\gamma\delta}g_{\delta\gamma}+e^{\frac{4}{3}(\phi-\varphi)}-1)
+e^{\frac{2}{3}(\varphi-\phi)}g_{\beta\alpha}=0 \ . \nonumber \\
\end{eqnarray}
Then in order to solve (\ref{eqphi}) we presume that  $\phi=\varphi$ and hence
(\ref{eqphi}) gives
$\gamma^{\alpha\beta}g_{\beta\alpha}=2$. Inserting this result into
(\ref{eqgammasim}) 	w finally obtain
\begin{equation}
\gamma_{\alpha\beta}=g_{\alpha\beta} \ .  
\end{equation}
Collecting all these results we find that the action (\ref{Sext}) reduces into standard
Nambu-Goto action for bosonic string
\begin{equation}
S=-T_{FS}\int \sqrt{-g} \ 
\end{equation}
which we wanted to show. 

The main goal of this section was to demonstrate how to perform dimensional reduction 
with auxiliary world-sheet metric. In the next section we apply this procedure for null
dimensional reduction of M-theory.

\section{Non-Relativistic String}\label{third}
In this section we proceed to the analysis of null dimensional reduction of M-theory and corresponding extended probe which is M2-brane.   To do this we presume that  the background
metric has light-like isometry
\cite{Julia:1994bs,Hartong:2015xda}, for recent extended analysis, see \cite{Hansen:2020pqs}
\begin{eqnarray}
& &ds^2=g_{MN}dx^Mdx^N=2T_\mu dx^\mu(du-M_\nu dx^\nu)+h_{\mu\nu}dx^\mu dx^\nu \ , \nonumber \\
& & g_{\mu u}=T_\mu \ , \quad g_{\mu\nu}\equiv \hH_{\mu\nu}=H_{\mu\nu}-M_\mu T_\nu-T_\mu M_\nu \ , \nonumber \\
\end{eqnarray}
where $\mu,\nu=0,1,\dots,9$ and where $u-$is light-like coordinate. 

In this section we consider M2-brane extended along light-like direction $u$. To do this we
use Polyakov form of M2-brane action (\ref{M2pol}) where we now presume that three dimensional world-volume metric has light-like isometry so that it can be written in the form
\begin{equation}
\gamma_{\alpha u}=\tau_\alpha \ , \quad  \gamma_{\alpha\beta}=
h_{\alpha\beta}-m_\alpha \tau_\beta-\tau_\alpha m_\beta \ . 
\end{equation}
Then we have
\begin{equation}
\det \gamma=\left|\begin{array}{cc}
0 & \tau_\beta \\
\tau_\alpha & h_{\alpha\beta}-m_\alpha \tau_\beta-
m_\beta \tau_\alpha \\ \end{array}\right|=
\left|\begin{array}{cc}
0 & \tau_\beta \\
\tau_\alpha & h_{\alpha\beta} \\ \end{array}\right|
=\det (-\tau_\alpha\tau_\beta+h_{\alpha\beta}) \ . 
\end{equation}
We further have an inverse metric
\begin{equation}
\gamma^{uu}=2\Phi \ , \quad \gamma^{u\alpha}=-\hv^\alpha \ , \quad 
\gamma^{\alpha\beta}=h^{\alpha\beta} \ , 
\end{equation}
where these fields obey the relation
\begin{equation}
h^{\alpha\beta}h_{\beta\gamma}-\tau_\gamma v^\alpha=\delta^\alpha_\gamma
\ , \quad 
\tau_\alpha v^\alpha=-1 \ , \quad  h_{\alpha\beta}v^\beta=h^{\alpha\beta}\tau_\beta=0 \ ,
\end{equation}
and where 
\begin{equation}
\hv^\alpha=v^\alpha -h^{\alpha\beta}m_\beta
\ , \quad 
\Phi=-v^\alpha m_\alpha+\frac{1}{2}h^{\alpha\beta}m_\alpha m_\beta \ . 
\end{equation}
Finally using the fact that the  induced metric has the form
\begin{equation}
g_{uu}=0 \ , \quad 
g_{u\alpha}=g_{u\mu}\partial_\alpha x^\mu=T_\alpha\ , 
\quad 
g_{\alpha\beta}=\hH_{\mu\nu}\partial_\alpha x^\mu \partial_\beta x^\nu
\end{equation}
we obtain following action
\begin{eqnarray}
& &S=-\frac{T_{M2}}{2}
\int du \int d^2\xi 
\sqrt{-\det (-\tau_\alpha \tau_\beta+h_{\alpha\beta})}
\times \nonumber \\
& &\times (-2\hv^\alpha T_{\alpha}+h^{\alpha\beta}\hat{H}_{\beta\alpha}-1) \ . 
\nonumber \\
\end{eqnarray}
Generally $\int du\rightarrow \infty$ since $u$ is non-compact coordinate. Then we can rescale M2-brane tension as $T_{M2}=\frac{T_{FS}}{\int du}$ so that
\begin{equation}
T_{FS}=T_{M2}\int du \ .
\end{equation}
 In summary, we have an action for non-relativistic string in the form
\begin{equation}\label{Sfinalns}
S=-\frac{T_{FS}}{2}\int d^2\xi\sqrt{-\det (-\tau_\alpha\tau_\beta+h_{\alpha\beta})}
(-2\hv^\alpha T_\alpha+h^{\alpha\beta}\hat{H}_{\beta\alpha}-1) \ . 
\end{equation}
 This is final form of non-relativistic string that is  defined by double  dimensional reduction of M2-brane along null direction.

The structure of the action 
(\ref{Sfinalns}) suggests that it has non-relativistic form on the world-sheet
due to the presence of two dimensional Newton-Cartan metric.  However  in order
to gain more insight into its structure let us solve equations  of motion for $h^{\alpha\beta},v^\alpha$ and $m_\alpha$. 
To do this we use the fact that the matrix $\bA_{\alpha\beta}=-\tau_\alpha\tau_\beta+h_{\alpha\beta}$ is non-singular and hence
it has an inverse matrix in the form 
 $\bA^{\alpha\beta}=-v^\alpha v^\beta+h^{\alpha\beta}$ so that
 \begin{equation}
 \det \bA_{\alpha\beta}=\frac{1}{\det \bA^{\alpha\beta}} \ .
 \end{equation}
 Then we can replace determinant $\det \bA_{\alpha\beta}$ with 
 $\frac{1}{\det \bA^{\alpha\beta}}$  and hence (\ref{Sfinalns})
 can be written in the form 
 \begin{equation}
 S=-\frac{T_{FS}}{2}\int d^2\xi \frac{1}{\sqrt{-\det \bA^{\alpha\beta}}}
 (-2\hv^\alpha T_\alpha+h^{\alpha\beta}\hH_{\beta\alpha}-1) \ . 
 \end{equation}
 Now we are ready to perform variation of this action with respect to 
 $v^\alpha,h^{\alpha\beta}$ and $m_\alpha$ treating them as independent variables. 
 Firstly we get equation of motion for $v^\alpha$
 \begin{eqnarray}\label{valpha}
 \tau_\alpha
 (-2\hv^\beta T_\beta+h^{\gamma\delta}\hH_{\delta\gamma}-1)
 - 2T_\alpha
=0 \ . \nonumber \\
  \end{eqnarray}
Further, equations of motion for $h^{\alpha\beta}$ have the form
\begin{eqnarray}\label{eqhab}
(-\tau_\alpha\tau_\beta+h_{\alpha\beta})(-2\hv^\gamma T_\gamma+
h^{\gamma\delta}\hH_{\delta\gamma}-1)-2(H_{\alpha\beta}+
(m_\alpha-M_\alpha)T_\beta+T_\alpha(m_\alpha-M_\beta))=0
\nonumber \\
\end{eqnarray}
and finally equations of motion for $m_\alpha$ has the form 
\begin{equation}
h^{\alpha\beta}T_\beta=0 \ . 
\end{equation}
This is very important equation that has natural solution when we identify
\begin{equation}\label{tauT}
\tau_\alpha=T_\alpha \ . 
\end{equation}
 Then the equation (\ref{valpha}) reduces into simple equation 
\begin{equation}
h^{\alpha\beta}H_{\beta\alpha}-1=0
\end{equation}
This equation has clearly solution in the form
\begin{equation}\label{hH}
h_{\alpha\beta}=H_{\alpha\beta}-k_\alpha \tau_\beta-\tau_\alpha k_\beta \ , 
\end{equation}
where $k_\alpha=K_\mu\partial_\alpha x^\mu$ is arbitrary two dimensional vector.
To see this let us consider the defining equation  
\begin{equation}
h^{\alpha\beta}h_{\beta\gamma}-v^\alpha \tau_\gamma=\delta^\alpha_\gamma \ . 
\end{equation}
Taking its trace we get
\begin{equation}
h^{\alpha\beta}h_{\beta\alpha}-v^\alpha \tau_\alpha=2
\end{equation}
that implies
\begin{equation}
h^{\alpha\beta}h_{\beta\alpha}=1
\end{equation}
using the fact that $v^\alpha \tau_\alpha=1$. Then we see that the equation
(\ref{eqhab}) has the form
\begin{equation}
-2\tau_\alpha\tau_\beta-2(m_\alpha-M_\alpha+k_\alpha)T_\beta-2T_\alpha(m_\beta-M_\beta+k_\beta)=0
\end{equation}
that has solution 
\begin{equation}
m_\alpha=M_\alpha+\frac{T_\alpha}{2}-k_\alpha \ . 
\end{equation}
Inserting (\ref{tauT}) and (\ref{hH}) into the action (\ref{Sfinalns}) we find
its square-root form 
\begin{equation}\label{Snon}
S=-T_{FS}\int d^2\xi \sqrt{-\det \bA_{\alpha\beta}}
=T_{FS}\int d^2\xi \sqrt{-\det (-T_\alpha T_\beta+H_{\alpha\beta}-k_\alpha T_\beta-
T_\alpha k_\beta)} \ . 
\end{equation}
This is the final form of the action for fundamental string in theory that arises through
null reduction of M-theory. We see that it has Nambu-Goto form  despite
of the fact that its Polyakov like form contains world-sheet metric that is non-relativistic. Further, we see that there is family of Newton-Cartan metrics 
corresponding to given null background due to the presence of the arbitrary vector $k_\alpha$. This is characteristic property of the Newton-Cartan background that is 
defined with the help of null reduction of higher dimensional background.

\section{Hamiltonian Formalism}\label{fourth}
In this section we find Hamiltonian formulation of the non-relativistic string 
with the action given in (\ref{Snon}).
 From (\ref{Snon}) we obtain following
conjugate momenta
\begin{equation}\label{pmu}
p_\mu=-T_{FS}\bA_{\mu\nu}\partial_\beta x^\nu \bA^{\beta 0}\sqrt{-\det \bA} \ , 
\end{equation}
where $\bA_{\mu\nu}=-T_\mu T_\nu+H_{\mu\nu}+k_\mu T_\nu+T_\mu k_\nu$. It is important to stress 
that the matrix $\bA_{\mu\nu}$ is non-singular and hence it has an inverse matrix equal to
\begin{equation}
\bA^{\mu\nu}=H^{\mu\nu}-\frac{1}{1+2\Phi}\hV^\mu\hV^\nu \ , 
\end{equation}
where
\begin{equation}
\hV^\mu=V^\mu-H^{\mu\nu}k_\nu \ , \quad \Phi=-k_\mu V^\mu+
\frac{1}{2}k_\mu H^{\mu\nu}k_\nu \ . 
\end{equation}
Then the bare Hamiltonian is equal to
\begin{equation}
\mH_B=p_\mu\partial_0x^\mu-\mL=-T_{FS}\bA_{0\beta}\bA^{\beta 0}+\sqrt{-\det \bA}=0
\end{equation}
as we should expect for theory that is invariant under two dimensional diffeomorphism. On the other hand we have two primary constraints that follow 
from definition of momenta $p_\mu$ (\ref{pmu})
\begin{equation}
\mH_S=p_\mu \partial_1
 x^\mu
 \approx 0 \  
\end{equation}
and also
\begin{equation}
\mH_T=p_\mu \bA^{\mu\nu}p_\nu+T_{FS}^2\bA_{\mu\nu}\partial_1 x^\mu
\partial_1 x^\nu\approx 0\ . 
\end{equation}
Then the total Hamiltonian is the sum of two primary constraints
\begin{equation}
H=\int d\xi^1(N^S\mH_S+N^T \mH_T) \ . 
\end{equation}
It would be simple exercise to show that $\mH_S,\mH_T$ are first class constraints 
which again reflect invariance of the action under two dimensional diffeomorphism. 
\section{Transverse dimensional reduction}\label{fifth}
Let us consider M2-brane that is transverse to the light-like direction. In this case the induced metric has the form 
\begin{equation}
\hg_{\bar{\alpha}\bar{\beta}}=\hH_{\balpha\bbeta}+T_{\balpha}\partial_{\bbeta}u+
\partial_{\balpha}u T_{\bbeta} \ 
\end{equation}
so that the  action (\ref{M2pol})
 has the form
\begin{eqnarray}\label{actM2}
S=-\frac{T_{M2}}{2}\int d^3\xi \sqrt{-\gamma}(\gamma^{\balpha\bbeta}(\hH_{\balpha\bbeta}
+\partial_{\balpha}u T_{\bbeta}+T_{\balpha}\partial_{\bbeta}u)-1) \ . 
\nonumber \\
\end{eqnarray}
We would like to introduce vector field dual to $\partial_{\balpha}u$ which is not possible directly in the action (\ref{actM2}) since there are no terms quadratic in 
$\partial_{\balpha}u$. In order to have terms quadratic in $\partial_{\balpha}u$
let us introduce two auxiliary fields $A_{\balpha},B_{\balpha}$ and consider following expression 
\footnote{Similar procedure can be found in 
	\cite{Bergshoeff:2019pij,Kluson:2018vfd}. }
\begin{equation}\label{auxAB}
\sqrt{-\gamma}(\gamma^{\balpha\bbeta}\partial_{\balpha}u\partial_{\bbeta}u+
A_{\balpha}\gamma^{\balpha\bbeta}\partial_{\bbeta}u+B_{\balpha}\gamma^{\balpha
\bbeta}\partial_{\bbeta}u+A_{\balpha}\gamma^{\balpha\bbeta}B_{\bbeta}) \ 
\end{equation}
that we add to the action (\ref{actM2}). Let us now consider  equations of motion  for $A_{\balpha},B_{\balpha}$ that follow from (\ref{auxAB})  
\begin{equation}
\partial_{\balpha}u+B_{\balpha}=0 \ , \quad 
\partial_{\balpha}u+A_{\alpha}=0 \ . 
\end{equation}
Inserting this result into (\ref{auxAB}) we see that this expression is zero and hence
original and extended action for M2-brane are the same when $A_{\balpha},B_{\balpha}$ are on-shell. However thanks to the presence of the term quadratic in derivatives of $u$ we
can dualize scalar $u$. To do this 
let us now introduce $Y_{\balpha}=\partial_{\balpha}u$. Clearly this vector obeys 
\begin{equation}\label{condY}
\epsilon^{\balpha\bbeta\bgamma}\partial_{\bbeta}Y_{\bgamma}=0 \ , 
\end{equation}
where $\epsilon^{\balpha\bbeta\bgamma}$ is totally antisymmetric symbol with 
$\epsilon^{012}=1$. 
Let us now interpret $Y_{\balpha}$ as an independent field  while condition (\ref{condY}) is replaced by following term in the  action 
\begin{equation}
T_{M2}\int d^3\xi \epsilon^{\balpha\bbeta\bgamma}
\partial_{\balpha}A_{\bbeta}Y_{\bgamma}=\frac{T_{M2}}{2}
\int d^3\xi \epsilon^{\balpha\bbeta\bgamma}(\partial_{\balpha}A_{\bbeta}-
\partial_{\bbeta}A_{\balpha})Y_{\bgamma} \ , 
\end{equation}
where now variation with respect to $A_{\bbeta}$ gives (\ref{condY}). In summary, the extended form of M2-brane action has the form
\begin{eqnarray}\label{SM2ext}
& &S=-\frac{T_{M2}}{2}\int d^3\xi \sqrt{-\gamma}(\gamma^{\balpha\bbeta}(\hH_{\balpha\bbeta}
+Y_{\balpha} T_{\bbeta}+T_{\balpha}Y_{\bbeta})
+\nonumber \\
&&+\gamma^{\balpha\bbeta}Y_{\balpha}Y_{\bbeta}+
A_{\balpha}\gamma^{\balpha\bbeta}Y_{\bbeta}+B_{\balpha}\gamma^{\balpha
	\bbeta}Y_{\bbeta}+A_{\balpha}\gamma^{\balpha\bbeta}B_{\bbeta}
-1)+\frac{T_{M2}}{2}\int d^3\xi  \epsilon^{\balpha\bbeta\bgamma}
F_{\balpha\bbeta}Y_{\bgamma} \ ,
\nonumber \\
\end{eqnarray}
where $F_{\balpha\bbeta}=\partial_{\balpha}A_{\bbeta}-\partial_{\bbeta}A_{\balpha}$.
Finally we eliminate $Y_{\balpha}$ by solving their equations of motion that follow from the action above and we get
\begin{eqnarray}
-\sqrt{-\gamma}\gamma^{\balpha\bbeta}(Y_{\bbeta}+T_{\bbeta}+\frac{1}{2}A_{\bbeta}+\frac{1}{2}B_{\bbeta})
+\frac{1}{2}\epsilon^{\balpha\bbeta\bgamma}
F_{\bbeta\bgamma}
=0 \ . 
\nonumber \\
\end{eqnarray}
From this equation we can express $Y_{\balpha}$ as 
\begin{equation}
Y_{\balpha}=\frac{1}{2\sqrt{-\gamma}}
\gamma_{\balpha\bdelta}\epsilon^{\bdelta \bbeta \bgamma}
F_{\bbeta\bgamma}-\frac{1}{2}(A_{\balpha}+B_{\balpha}+2T_{\balpha}) \ . 
\end{equation}
Then inserting this result into the action (\ref{SM2ext}) 
we obtain 
\begin{eqnarray}
& &S=-\frac{T_{M2}}{2}
\int d^3\xi \sqrt{-\gamma}(\gamma^{\balpha\bbeta}\hH_{\bbeta\balpha}
-\frac{1}{2}F_{\balpha\bbeta}\gamma^{\bbeta\bbeta'}
F_{\bbeta'\balpha'}
\gamma^{\balpha'\balpha}
-\nonumber \\
& &-\frac{1}{4}
(A-B)_{\balpha}\gamma^{\balpha\bbeta}(A-B)_{\bbeta}-(A+B)_{\balpha}\gamma^{\balpha\bbeta}T_{\bbeta}
-T_{\balpha}\gamma^{\balpha\bbeta}T_{\bbeta}-1)
\nonumber \\
& &-\frac{T_{M2}}{4}
\int d^3\xi \epsilon^{\balpha\bbeta\bgamma}
F_{\balpha\bbeta}(A_{\bgamma}+B_{\bgamma}+2T_{\bgamma}) \  \nonumber \\
\end{eqnarray}
using the fact that 
\begin{eqnarray}
\frac{1}{(\sqrt{-\gamma})^2}F_{\bar{\gamma}\bbeta}\epsilon^{\bgamma\bbeta \bdelta}\gamma_{\bdelta\balpha}\epsilon^{\balpha\bar{\omega}\bar{\sigma}}
F_{\bar{\omega} \bar{\sigma}}=-2F_{\balpha\bbeta}\gamma^{\balpha\bar{\omega}}\gamma^{\bbeta\bdelta}F_{
	\bar{\omega}\bdelta} \ . 
\nonumber \\
\end{eqnarray}
As the next step we  introduce two fields $X_{\balpha}=A_{\balpha}-B_{\balpha}$ and $2Z_{\balpha}=A_{\balpha}+B_{\balpha}$. We see that  the equations of motion for $X_{\balpha}$ implies
that $X_{\balpha}=0$. Then the action has the form 
\begin{eqnarray}
& &S=-\frac{T_{M2}}{2}
\int d^3\xi \sqrt{-\gamma}(\gamma^{\balpha\bbeta}(-T_{\balpha}T_{\bbeta}+H_{\balpha\bbeta}-(M_{\balpha}+Z_{\balpha})T_{\bbeta}+
T_{\balpha}(M_{\bbeta}+Z_{\bbeta}))-\nonumber \\
& &-\frac{1}{2}F_{\balpha\bbeta}\gamma^{\bbeta\bbeta'}
F_{\bbeta'\balpha'}
\gamma^{\balpha'\balpha}-1)
-\frac{T_{M2}}{2}
\int d^3\xi \epsilon^{\balpha\bbeta\bgamma}
F_{\balpha\bbeta}(Z_{\bgamma}+T_{\bgamma}) \ . \nonumber \\
\end{eqnarray}
Note that this is similar form of the action as was derived in 
\cite{Townsend:1995af} in case of standard dimensional reduction of M-theory.

Finally we should solve equations of motion for $\gamma^{\balpha\bbeta}$ that follow from the action above
\begin{eqnarray}
-\frac{1}{2}\gamma_{\balpha\bbeta}
	(\gamma^{\bgamma\bdelta}\mathbf{H}_{\bdelta\bgamma}
-\frac{1}{2}F_{\balpha\bbeta}\gamma^{\bbeta\bbeta'}
F_{\bbeta'\balpha'}
\gamma^{\balpha'\balpha}-1)	+\mathbf{H}_{\balpha\bbeta}
-F_{\bgamma\balpha}F_{\bbeta \bdelta}\gamma^{\bdelta\bgamma}=0 \ , 
\nonumber \\
\end{eqnarray}
where we defined
\begin{equation}
\mathbf{H}_{\balpha\bbeta}=-T_{\balpha}T_{\bbeta}+H_{\balpha\bbeta}
-(M_{\balpha}+Z_{\balpha})T_{\bbeta}-(M_{\bbeta}+Z_{\bbeta})T_{\balpha} \ . 
\end{equation}
In the leading order approximation it has solution \cite{Townsend:1995af} $\gamma_{\balpha\bbeta}=\mathbf{H}_{\balpha\bbeta}$ so that in the leading order approximation the action has the form 
\begin{eqnarray}\label{SM2final}
& &S=-T_{M2}\int d^3\xi \sqrt{-\det \mathbf{H}_{\balpha\bbeta}}
(1-\frac{1}{4}
F_{\balpha\balpha'}
\mathbf{H}^{\balpha' \bdelta}F_{\bdelta \bgamma}\mathbf{H}^{\bgamma\balpha}
)=\nonumber \\
& &\approx-T_{M2}
\int d^3\xi \sqrt{-\det \mathbf{H}_{\balpha\bbeta}
(1-\frac{1}{2}
F_{\balpha\balpha'}
\mathbf{H}^{\balpha' \bdelta}F_{\bdelta \bgamma}\mathbf{H}^{\bgamma\balpha}
)}=\nonumber \\
& &=
-T_{M2}
\int d^3\xi \sqrt{-\det (\mathbf{H}_{\balpha\bbeta}+F_{\balpha\bbeta}) \ , 
} \nonumber \\
\end{eqnarray}	
where in the last step we used relations that hold in three dimensions
\begin{equation}
\det (\mathbf{H}_{\balpha\bbeta}+F_{\balpha\bbeta})=\det \mathbf{H}_{\balpha\bbeta}
\det (\delta_{\balpha}^{\bbeta}+F_{\balpha\bgamma}\mathbf{H}^{\bgamma\bbeta})=\det \mathbf{H}_{\balpha\bbeta}\left(
1-\frac{1}{2}F_{\balpha\balpha'}
\mathbf{H}^{\balpha' \bdelta}F_{\bdelta \bgamma}\mathbf{H}^{\bgamma\balpha}\right) \ .
\end{equation}
The action (\ref{SM2final}) is final form of D2-brane action in non-relativistic string
theory that is defined by null dimensional reduction of M-theory. We again observe an interesting fact that given M-theory background with null isometry is mapped into 
family of Newton-Cartan  backgrounds that differ by redefinition of $M_\mu$. In particular, choosing $Z_{\mu}=-M_\mu$ we get the action
\begin{equation}
S=-T_{M2}\int d^3\xi \sqrt{-\det (-T_{\balpha}T_{\bbeta}+H_{\balpha\bbeta}+F_{\balpha\bbeta})}
-\frac{T_{M2}}{2}\int d^3\xi \epsilon^{\balpha\bbeta\bgamma}F_{\balpha\bbeta}
(T_{\bgamma}-M_{\bgamma}) \ . 
\end{equation}
In summary, we found D2-brane action in the theory that is defined by null dimensional reduction of M-theory. It would be certainly interesting to extend this analysis to the case of M5-brane and its null dimensional reduction and study how
the procedure suggested in this paper could be related to the recent discussion of null dimensional reduction of M5-branes presented in 
\cite{Lambert:2020scy}.

{\bf Acknowledgement:}
\\
	This work 
	is supported by the grant “Integrable Deformations”
	(GA20-04800S) from the Czech Science Foundation
	(GACR).


\end{document}